\documentstyle[12pt]{article}
\begin{document}
\newcommand{\half}{{\textstyle\frac{1}{2}}}
\newsavebox{\uuunit}
\sbox{\uuunit}
    {\setlength{\unitlength}{0.825em}
     \begin{picture}(0.6,0.7)
        \thinlines
        \put(0,0){\line(1,0){0.5}}
        \put(0.15,0){\line(0,1){0.7}}
        \put(0.35,0){\line(0,1){0.8}}
       \multiput(0.3,0.8)(-0.04,-0.02){12}{\rule{0.5pt}{0.5pt}}
     \end {picture}}
\newcommand {\unity}{\mathord{\!\usebox{\uuunit}}}
\newcommand{\eentwee}{-1 \leftrightarrow 2}
\newcommand{\een}{{(1)}}
\newcommand{\twee}{{(2)}}
\newcommand{\drie}{{(3)}}
\newcommand  {\Rbar} {{\mbox{\rm$\mbox{I}\!\mbox{R}$}}}
\newcommand  {\Hbar} {{\mbox{\rm$\mbox{I}\!\mbox{H}$}}}
\newcommand {\Cbar}
    {\mathord{\setlength{\unitlength}{1em}
     \begin{picture}(0.6,0.7)(-0.1,0)
        \put(-0.1,0){\rm C}
        \thicklines
        \put(0.2,0.05){\line(0,1){0.55}}
     \end {picture}}}
\newcommand{\cL}{{\cal L}}
\newcommand{\cB}{{\cal B}}
\newcommand{\cM}{{\cal M}}
\newcommand{\cN}{{\cal N}}
\newcommand{\cZ}{{\cal Z}}
\newcommand{\Al}{Alekseevski\v{\i}}
\newcommand{\eqn}[1]{eq.(\ref{#1})}

\renewcommand{\section}[1]{\addtocounter{section}{1}
\vspace{5mm} \par \noindent
  {\bf \thesection . #1}\setcounter{subsection}{0}
  \par
   \vspace{2mm} } 
\newcommand{\sectionsub}[1]{\addtocounter{section}{1}
\vspace{5mm} \par \noindent
  {\bf \thesection . #1}\setcounter{subsection}{0}\par}
\renewcommand{\subsection}[1]{\addtocounter{subsection}{1}
\vspace{2.5mm}\par\noindent {\em \thesubsection . #1}\par
 \vspace{0.5mm} }
\renewcommand{\thebibliography}[1]{ {\vspace{5mm}\par \noindent{\bf
References}\par \vspace{2mm}}
\list
 {\arabic{enumi}.}{\settowidth\labelwidth{[#1]}\leftmargin\labelwidth
 \advance\leftmargin\labelsep\addtolength{\topsep}{-4em}
 \usecounter{enumi}}
 \def\newblock{\hskip .11em plus .33em minus .07em}
 \sloppy\clubpenalty4000\widowpenalty4000
 \sfcode`\.=1000\relax \setlength{\itemsep}{-0.4em} }

\begin{titlepage}
\begin{flushright} THE-93/24\\ KUL-TF-93/43\\ hep-th@xxx/9310067
\end{flushright}
\vfill
\begin{center}
{\large\bf HIDDEN SYMMETRIES, SPECIAL GEOMETRY AND QUATERNIONIC
MANIFOLDS${}^\dagger$ }   \\
\vskip 7.mm
{B. de Wit }\\
\vskip 0.1cm
{\em Institute for Theoretical Physics} \\
{\em Utrecht University}\\
{\em Princetonplein 5, 3508 TA Utrecht, The Netherlands} \\[5mm]
 A. Van Proeyen \\
\vskip 1mm
{\em Instituut voor theoretische fysica}\\
{\em Universiteit Leuven, B-3001 Leuven, Belgium}
\end{center}
\vfill

\begin{center}
{\bf ABSTRACT}
\end{center}
\begin{quote}
The moduli space of the Calabi-Yau three-folds, which play a role
as superstring ground states, exhibits the same {\em special
geometry} that is known from nonlinear sigma models in $N=2$
supergravity theories. We discuss the symmetry structure of
special real, complex and quaternionic spaces. Maps between these
spaces are  implemented via dimensional reduction. We analyze the
emergence of {\em extra} and {\em hidden} symmetries. This
analysis is then applied to homogeneous special spaces and the
implications for the classification of homogeneous quaternionic
spaces are discussed.

\vfill      \hrule width 5.cm
\vskip 2.mm
{\small\small
\noindent $^\dagger$ Invited talk given at the Journ\'ees
Relativistes '93, Brussels, 5-7 April 1993; to be
published in the proceedings.}
\end{quote}
\begin{flushleft}
September 1993
\end{flushleft}
\end{titlepage}

\vspace{4mm}
\begin{center}
{\bf HIDDEN SYMMETRIES, SPECIAL GEOMETRY AND QUATERNIONIC MANIFOLDS }
\vspace{1.4cm}

B.~DE~WIT \\
{\em Institute for Theoretical Physics, Utrecht University} \\
{\em Princetonplein 5, 3508 TA Utrecht, The Netherlands} \\
\vspace{5mm}
A. VAN~PROEYEN\\
{\em Instituut voor theoretische fysica}\\
{\em Universiteit Leuven, B-3001 Leuven, Belgium} \\
\end{center}
\centerline{ABSTRACT}
\vspace{- 4 mm}  
\begin{quote}\small
The moduli space of the Calabi-Yau three-folds, which play a role
as superstring ground states, exhibits the same {\em special
geometry} that is known from nonlinear sigma models in $N=2$ supergravity
theories. We discuss the symmetry structure of special real,
complex and quaternionic spaces. Maps between these spaces are
implemented via dimensional reduction. We analyze the emergence of
{\em extra} and {\em hidden} symmetries. This analysis is then
applied to
homogeneous special spaces and the implications for the
classification of homogeneous quaternionic spaces are discussed.
\end{quote}
\addtocounter{section}{1}
\par \noindent
  {\bf \thesection . Introduction}
  \par
   \vspace{2mm} 

\noindent
Upon dimensional reduction a field theory may
exhibit certain unexpected symmetries. These symmetries are often called
{\em hidden} symmetries.  Some of them are clearly related
to the symmetries of the original higher-dimensional theory,
while for others there is no obvious explanation.
A simple and well-known example of this phenomenon is Einstein
gravity, which leads to a nonlinear sigma model coupled to
gravity upon reduction to three space-time dimensions. This sigma
model is associated with the symmetric space $SO(2,1)/SO(2)$, which is
invariant under an $SO(2,1)$ group of isometries.

Hidden symmetries play, for instance, a role in Kaluza-Klein theories, in
the study of lower-dimensional solutions of the Einstein equation,
and in supergravity. In the latter dimensional
reduction has been used as a convenient method for obtaining
information about these theories in various dimensions.
This often led to the discovery of new structures
and unexpected connections with intriguing mathematics. The
latter is especially the case when the dimensional reduction
preserves supersymmetry. Supersymmetry poses restrictions on the
symmetries of the matter sector of the lower-dimensional theory.
In the context of this talk we are interested in
the K\"ahler geometry associated with $N=2$ vector multiplets
coupled to supergravity (supergravity invariant under two
independent local supersymmetries) \cite{dWVP}. This geometry is
called {\em special geometry} \cite{special}.  New impetus for
studying hidden symmetries and special geometry came from
superstring compactifications on Calabi-Yau manifolds. Again the
method of dimensional reduction turns out to play a useful role
here, which motivated us to try and explain the structure of
hidden symmetries in the context of special geometry. Somewhat
surprisingly, this study led to implications on the
classification of homogeneous quaternionic spaces.

In order to clarify the connections between special geometry,
hidden symmetries, Calabi-Yau manifolds and superstrings, let us
first explain a few facts concerning superstring compactifications.
Because the superstring lives in ten space-time dimensions,
realistic theories require six of the spatial dimensions
to be compactified. The corresponding superstring ground states
can be described in terms of conformal field theories on the
superstring worldsheet. In many cases the relation between such
a conformal field theory and the compactification of the six
coordinates is obvious, but there exist conformal theories
without a space-time interpretation. In order to obtain realistic
low-energy field
theories one requires the compactification to be
supersymmetric \cite{GrosCand}\footnote{The supersymmetry will
eventually be broken at a distance scale much larger than the
compactification scale. The precise mechanism for supersymmetry
breaking is as yet not clear.}.
Supersymmetry implies that the compactified dimensions should
constitute a so-called Calabi-Yau three-fold (a compact K\"ahler
manifold of vanishing Chern class and complex dimension three;
these spaces have a unique Ricci-flat metric). In terms of
conformal field theories, the class of $(2,2)$
superconformal field theories with central charge $c=9$ is
relevant, which contains the
Calabi-Yau three-folds, but possibly also other solutions without
a corresponding space-time interpretation. In the context of this
work we ignore this aspect and generically denote this class of
ground states as Calabi-Yau manifolds. It is important to note
that these manifolds can serve as a ground state for each of the
three types of superstrings: \vspace{-3mm}
\begin{itemize}
\item For the so-called {\em heterotic string} they give rise to
low-energy effective theories that exhibit space-time $N=1$
supersymmetry. All phenomenologically viable models belong to
this class.  \vspace{-3mm}
\item For the so-called IIA and IIB superstrings they give rise to
theories with space-time $N=2$ supersymmetry. \vspace{-3mm}
\end{itemize}
The type-II superstrings have a more restrictive symmetry structure.
They are {\em not}
phenomenologically viable, but the fact that they can be
compactified on the same Calabi-Yau manifold implies that many of
their systematic features carry over to the compactifications of
the heterotic string. Our strategy is to make maximal use of this
fact (following \cite{Seiberg}) and base our study on $N=2$
space-time supersymmetry, in spite of the fact that realistic
low-energy effective theories exhibit only $N=1$ supersymmetry.

There exists a huge variety of string ground states
corresponding to Calabi-Yau spaces, parametrized by
parameters called {\em moduli}. The moduli thus correspond to the
independent deformations of the Ricci-flat metric up to
reparametrizations (see, for instance, \cite{Cand}). The
mixed deformations (i.e., related to the components of the metric
with one holomorphic and one anti-holomorphic index) are related
to the real harmonic $(1,1)$ forms, while the
pure (anti-)holomorphic deformations are related to the complex
harmonic $(2,1)$ forms on the Calabi-Yau space. For given
topology these spaces are thus described by $h_{11}$ real and
$h_{12}$ complex parameters, where the Hodge numbers $h_{pq}$
define the number of independent
harmonic $(p,q)$ forms. The $(1,1)$ moduli parametrize the
deformations of the K\"ahler class, which
characterize the size of the Calabi-Yau manifold. The $(2,1)$
moduli parametrize the deformations of the complex structure and
characterize the shape of the Calabi-Yau manifold.

\vspace{1mm}
\noindent
\begin{minipage}{8.7cm} \setlength{\parindent}{6mm}  The Hodge
diamond for a Calabi-Yau manifold is shown on the right. The only
Hodge numbers that are not uniquely determined are
$h_{12}=h_{21}$ and $h_{11}=h_{22}$. Knowing the numbers of
independent harmonic forms allows us to count the number of
massless Kaluza-Klein modes for each of
the various fields that occur in supergravity in ten dimensions,
compactified on a Calabi-Yau manifold (see, e.g. \cite{BoCa}).
For instance, the
ten-dimensional metric decomposes into a four-dimensional tensor,
a tensor with mixed indices, and a six-dimensional
tensor. \hfill The
\end{minipage}
\hfill
\begin{minipage}{5.8cm}
\setlength{\unitlength}{1mm}
\begin{picture}(60,60)
\put(30,54){1}
\put(23,46){0} \put(37,46){0}
\put(16,38){0} \put(29,38){$h_{11}$} \put(44,38){0}
\put(9,30){1}  \put(22,30){$h_{12}$} \put(36,30){$h_{12}$} \put(50,30){1}
\put(16,22){0} \put(29,22){$h_{11}$} \put(44,22){0}
\put(23,14){0} \put(37,14){0}
\put(30,6){1}
\end{picture}
\end{minipage}

\vspace{.9mm}\noindent
four-dimensional tensor is associated with the massless spin-2
graviton. The mixed components could in principle describe spin-1 states,
but as there are no $(1,0)$ or $(0,1)$ harmonic forms, these are
absent. Finally, the massless spin-0 states associated with the
pure six-dimensional components are precisely related to the
modular parameters of the Calabi-Yau space. Consequently, they
correspond to $h_{11}$
real and $h_{12}$ complex states. Using some elementary knowledge
of the various field equations, we can thus generally analyze the metric
$g_{MN}$, a scalar field $\phi$, a vector gauge field $A_M$,
antisymmetric tensor gauge fields $A_{MN}$ and $A_{MNP}$ and a
four-rank antisymmetric gauge field $A_{MNPQ}$ with
(anti)selfdual field strength and determine the number of
massless Kaluza-Klein states. With these results, summarized in
table \ref{KKstates}, it is easy to count the
number of bosonic massless states that emerge in the
compactification of IIA and IIB supergravity on a Calabi-Yau
manifold:
\begin{eqnarray}
\mbox{nonchiral IIA SG : }
\begin{array}{ll}
\left.\begin{array}{l}
h_{11} + 1 \mbox{ spin-1}  \\
h_{11} \mbox{ complex spin-0}\hspace{9mm}
\end{array}  \right\}
& h_{11} \mbox{ vector supermultiplets} \\[1mm]
\hspace{2mm} h_{12}+1 \mbox{ quaternionic spin-0} & h_{12}+1
\mbox{ scalar supermultiplets}
\end{array} \\[2mm]
\mbox{chiral IIB SG : }
\begin{array}{ll}
\left.\begin{array}{l}
h_{12} + 1 \mbox{ spin-1}  \\
h_{12} \mbox{ complex spin-0} \hspace{9mm}
\end{array}  \right\}
& h_{12} \mbox{ vector supermultiplets} \\[1mm]
\hspace{2mm} h_{11}+1 \mbox{ quaternionic spin-0} & h_{11}+1
\mbox{ scalar supermultiplets}
\end{array}
\end{eqnarray}
On the right we grouped the fields into $N=2$ vector and
scalar supermultiplets in four space-time dimensions (note that
the extra vector states belong to the graviphoton of the
supergravity multiplet). The bosonic
fields of a vector multiplet
are a vector gauge field and a complex scalar field. These
complex scalar fields are described by a nonlinear sigma model
associated with a K\"ahler space. Each vector multiplet thus
gives rise to one spin-1 and two spin-0 states. The bosonic
fields of a scalar multiplet are just four scalar fields,
described by a nonlinear sigma model associated with a
quaternionic space. The bosonic fields of a scalar
multiplet give rise to four spin-0 states.

\begin{table}[tf]
\begin{center}
\begin{tabular}{||cc||c||c|c|c||}
\hline
A & B & field & spin-2 & spin-1 & spin-0  \\
\hline &&&&&\\[-3mm]
1 & 1 & $g_{MN}$  & 1 & 0 & $h_{11}$ real $+$ $h_{12}$ complex \\[2mm]
1 & 2 & $\phi$    & 0 & 0 &  1    \\[2mm]
1 & 0 & $A_M$     & 0 & 1 &  0    \\[2mm]
1 & 2 & $A_{MN}$  & 0 & 0 & $(h_{11}+1)$ real \\[2mm]
1 & 0 & $A_{MNP}$ & 0 & $h_{11}$  & $(h_{12}+1)$ complex  \\[2mm]
0 & 1 & $\big[A_{MNPQ}\big]_\pm$ & 0 & $ h_{12}+1$ & $h_{11}$ real \\[2mm]
\hline
\end{tabular}
\end{center}
\caption{Massless Kaluza-Klein modes associated with various
fields in ten dimensions, compactified on a Calabi-Yau space. The
first two columns specify the number of these fields contained in IIA or
IIB supergravity in ten space-time dimensions. }
\label{KKstates}
\end{table}

There is a subtle relation between the moduli space of
Calabi-Yau manifolds and these K\"ahlerian and quaternionic sigma
models. In the effective low-energy theory corresponding to a
string compactification, the sigma-model fields have no
potential. Therefore their vacuum-expectation values are
undetermined and parametrize the (classical) ground states states
of this field theory, up to certain equivalence transformations.
The metric of
the non-linear sigma model (in four space-time dimensions) is
therefore related to the metric on moduli space \cite{Seiberg}.
This implies that the moduli space of the superstring
groundstates associated with Calabi-Yau spaces must exhibit
special geometry. Here we extend the notion of special
geometry to the quaternionic manifolds that emerge
in this context. A more precise definition will be given shortly.

An intriguing  feature is the existence of {\em mirror pairs}
\cite{mirror}. Under the interchange of $h_{11}$ and $h_{12}$ the
Hodge diamond changes into its mirror image by reflecting about a
diagonal. Manifolds related by such a reflection form a mirror
pair of two topologically {\em different} spaces. Nevertheless
they correspond to the same conformal field theory (except
that some of the $U(1)$ charge assignments have been reversed
when identifying the geometrical objects). From (1) and (2) it is
clear that this change is related to a change of chirality of the
corresponding supergravity theory, converting IIA and IIB
supergravity. This suggests that there must exist a map between
$n$-dimensional K\"ahler manifolds and $(n+1)$-dimensional
quaternionic manifolds. As was demonstrated by \cite{CecFerGir},
this map can be induced at the level of $N=2$ supergravity by
dimensional reduction from four to three dimensions. Under this
reduction the bosonic degrees of freedom associated with the
vector multiplets whose scalar fields parametrize a special
K\"ahler manifold are all converted into scalar fields, which
parametrize a quaternionic manifold. The quaternionic
manifolds that emerge through this so-called {\bf c}~map
are called {\em special quaternionic} manifolds.

Likewise we can define {\em special real} geometry as the
geometry associated with the sigma models that arise in $N=2$
supergravity coupled to vector multiplets in five space-time
dimensions. Upon dimensional reduction these real manifolds give rise
to a subclass of the special K\"ahler manifolds. The
corresponding map is called the {\bf r}~map. We should
emphasize that supersymmetry is the crucial ingredient in this
construction. As dimensional reduction preserves supersymmetry,
the manifolds that emerge under these maps must satisfy the
restrictions appropriate to $N=2$ supersymmetric matter. That means that
the scalar fields of vector multiplets parametrize a real
manifold in five \cite{GuSiTo} and a K\"ahler manifold in four
dimensions \cite{dWVP}, while scalar multiplets
parametrize quaternionic manifolds in four \cite{BagWit} and in
three dimensions \cite{dWTNic}. This ensures the existence of the {\bf
r} and {\bf c} maps, which act according to
\begin{equation}
\Rbar_{n-1} \stackrel{\bf r}{\longrightarrow} \Cbar_n \ ,\qquad
\Cbar_n \stackrel{\bf c}{\longrightarrow} \Hbar_{n+1}\ ,
\end{equation}
where $n-1$, $n$ and $n+1$ denote the real, complex and
quaternionic dimension of the real, K\"ahler and quaternionic
spaces, respectively.

The $\bf r$ and $\bf c$ maps are convenient tools for studying
the symmetry structure of the various special geometries. It is
here that the hidden symmetries enter. As we discuss below, we
make a distinction between {\em extra} symmetries that can be
understood from the invariances of the
higher-dimensional theory and {\em hidden} symmetries that
have no immediate explanation. The maps are
particularly useful for studying homogeneous special spaces,
because a homogeneous
space in the image of one of these maps must originate from a
theory in which the combined transformations of scalar and vector
fields act transitively on the scalars and vice versa. Motivated
by this relation we present a classification of the real
homogeneous spaces, whose image under the {\bf r} and {\bf
c}$\scriptstyle\circ${\bf r} map leads to corresponding
homogeneous K\"ahler and
quaternionic spaces. The latter are then confronted with
\Al's classification of the normal quaternionic spaces
\cite{Aleks} and their related K\"ahler spaces discussed in
\cite{Cecotti}. Our analysis allows a complete
determination of the isometry and isotropy groups of the
various spaces \cite{dWVPVan}.

\section{Special K\"ahler geometry}
\noindent
As discussed above, special K\"ahler geometry is the geometry that
arises when coupling $n$ vector supermultiplets to $N\!=\!2$
supergravity in
$d=4$ space-time dimensions. A characteristic feature is that
the geometry is encoded in a single holomorphic function $F(X)$ of
$n+1$ complex parameters $X^I$, where $I= 0, 1, \ldots,n$, which
is homogeneous of second degree. Therefore it satisfies
identities such as $F=\frac{1}{2}F_IX^I$, $F_I= F_{IJ}X^J$,
$X^IF_{IJK}=0$, where the subscripts $I,J,\dots$ denote
differentiation with respect to $X^I$, $X^J$, etc.
Furthermore it is convenient to define the tensor $N_{IJ}\equiv
{\frac{1}{2}}{\rm Re}\, F_{IJ}$. The $X^I$ are
scalar fields, which appear in the Lagrangian according to \cite{dWVP}
\begin{equation}
{\cal L}\propto \bigg(N_{IJ}-\frac{(N\bar X)_I\,(NX)_J}{\bar XN X}
\bigg) \partial_\mu X^I\,\partial^\mu \bar X^J\,,
\label{4dLagr}
\end{equation}
where we used an obvious notation where $(NX)_I=
N_{IJ}X^J$, $\bar XN X=\bar X^IN_{IJ}X^J$, etc.. Although the metric in
(\ref{4dLagr}) is degenerate, this poses no problems because the
overall scale of the fields is fixed by the condition $N_{IJ}\,
\bar X{}^I X^J=$~constant. Furthermore the
Lagrangian does not depend on the overall phase of the fields.
(In other words (\ref{4dLagr}) is invariant under local phase
transformations.)  Therefore it depends only on $n$ complex
fields $z^A$, which can be used to parametrize the $X^I$.
Introducing $n+1$ unrestricted functions $X^I(z)$, we
substitute everywhere for $X^I$,\footnote{The so-called {\em
special} coordinates are defined by $X^0(z)=1$, $X^A(z)=z^A$.}
\begin{equation}
X^I \longrightarrow {X^I(z) \over \sqrt{N_{KL}\,\bar X^K(z)\,
X{}^L(\bar z)}} \,.
\end{equation}
The kinetic term for the fields $z^A$ then
takes the form of a non-linear sigma model corresponding to a
K\"ahler manifold. Its K\"ahler potential is equal
to\footnote{The connection field associated with the local phase
transformations takes the form $A_\mu = \mbox{Im}\big(\partial_\mu
z^A\, \partial_AK(z,\bar z)\big)$. This quantity acts
also as a connection for K\"ahler transformations. Under these
transformations the K\"ahler potential is changed according to
$K(z,\bar z)\to K(z,\bar z)+f(z) + \bar f(\bar z)$, so that the
metric remains invariant, while $A_\mu$ changes
into $A_\mu +\partial_\mu \big(\mbox{Im }f(z)\big)$. }
\begin{equation}
K(z,\bar z ) =  \ln N_{IJ}\,X^I(z) \,\bar X{}^J(\bar z) =
\ln {\textstyle{1\over 2}}\,\mbox{Re } F_I\big(X(z)\big) \, \bar
X{}^I(z) \, ,
\end{equation}
and the sigma model metric is given by
\begin{equation}
g_{A\bar B} = \frac{\partial^2 K(z,\bar z)}{\partial z^A\,
\partial\bar z^B} \,.
\end{equation}
The curvature tensor corresponding to this metric equals
\cite{BEC}
\begin{equation}
R^A{}_{BC}{}^D = -2 \delta_{(B}^A\,\delta_{C)}^D - e^{-2K} \,
Q_{BCE}\,\bar Q^{EAD}\, ,
\end{equation}
where
\begin{equation}
Q_{ABC}(z) \equiv \textstyle{\frac{1}{4}} F_{IJK}\big(X(z)\big) \,
{\displaystyle{\partial X^I(z)\over \partial z^A} {\partial X^J(z)\over
\partial z^B} {\partial X^K(z)\over \partial z^C}}  , \quad
\bar Q^{ABC} = g^{\bar D A}g^{\bar E B}g^{\bar F C} \bar Q_{DEF} \,.
\end{equation}

Two different functions $F(X)$ can correspond to supergravity
theories with equivalent
equations of motion. Some of the equivalence transformations
involve duality transformations on the electric and magnetic
field strengths of the gauge fields. The reparametrizations
constitute a representation of $Sp(2n+2,\Rbar)$. For the
corresponding Calabi-Yau manifolds these symplectic
reparametrizations are naturally induced on the periods by changes
in the corresponding cohomology basis (see, e.g.
\cite{Cand,mirror}). A subclass of the reparametrizations may
correspond to an invariance of the equations of motions. For the
scalar fields these transformations constitute isometries of the
K\"ahler manifold.

\section{Special real geometry}
\noindent
Special real geometry is related to the functions
\begin{equation}
F(X)= i d_{ABC} {X^AX^BX^C\over X^0} , \label{Fd}
\end{equation}
with $d_{ABC}$ a symmetric real tensor. This tensor defines a
Maxwell-Einstein supergravity theory in five space-time
dimensions \cite{GuSiTo}, which contains $n$ real scalar fields
$h^A$ and $n$ vector fields $A_\mu^A$ (one of them corresponding
to the graviphoton). The relevant bosonic Lagrangian is
\begin{eqnarray}
e^{-1}{\cal L} &=& -{\textstyle{1\over 2}} R -{\textstyle{3\over
2}}\,(d\,h)_{AB} \,\partial_\mu h^A\,\partial^\mu h^B  \nonumber\\
&& +{\textstyle{1\over 2}}\,\left( 6(d\,h)_{AB} - 9 (d\,h h)_{A}\,
(d\,hh)_B \right) F^A_{\mu\nu}(A) \,F^{B\,\mu\nu}(A)
\nonumber\\
&& + e^{-1} i\epsilon^{\mu\nu\rho\sigma\lambda} d_{ABC} \,
F^A_{\mu\nu}(A)\,F^B_{\rho\sigma}(A)\, A^C_\lambda ,
\label{5dLagr}
\end{eqnarray}
where $(d\,h)_{AB}=d_{ABC}h^C$, $(d\,hh)_{A}=d_{ABC}h^Bh^C$, $R$
is the Ricci scalar and $F_{\mu\nu}^A(A)$ are the abelian
field-strength tensors corresponding to $A_\mu^A$. The scalar fields
$h^A$ are subject to the condition
\begin{equation}
d_{ABC}\,h^A h^B h^C =1.
\end{equation}
and parametrize an $(n-1)$-dimensional special real space whose
metric follows from the sigma-model metric in (\ref{5dLagr}).
The Lagrangian (\ref{5dLagr}) is invariant under linear
transformations of the fields
\begin{equation}
h^A \to \tilde B^A_{\;B}\, h^B\,, \qquad  A^A_\mu \to \tilde
B^A_{\;B}\,A^B_\mu\,, \label{Btrans}
\end{equation}
that leave the tensor $d_{ABC}$ invariant. These transformations
induce isometries on the special real space.

After reduction to four dimensions, the Lagrangian (\ref{5dLagr})
leads to a nonlinear sigma model associated with a special
K\"ahler space corresponding to (\ref{Fd}); the imaginary parts of the
four-dimensional scalar fields $z^A$ (in special coordinates) originate
from the fifth component of the gauge fields $A^A_\mu$, while
their real part corresponds to the $n-1$
independent fields $h^A$ and the component $g_{55}$
of the metric. The $n+1$ gauge fields in four dimensions are related
to the $n$ gauge fields and the off-diagonal
components $g_{\mu 5}$ of the metric in five dimensions. The
conversion of the bosonic fields is summarized in
table~2. Clearly in four dimensions we have $n+1$
extra scalar fields.

\begin{table}[t]
\begin{minipage}{5cm} Table 2: Decomposition of the $d=5$ bosonic
fields into $d=4$ fields.
\end{minipage}\hfill
\begin{minipage}{8cm}
\begin{tabular}{|r||c|c|c||}\hline
$d=5$ & metric  & vectors & scalars \\ \hline \hline
metric & $1$  & $1$ & $1$  \\
$n$ vectors & $0$ & $n$ & $n$ \\
$n-1$ scalars & $0$ & $0$ & $n-1$ \\ \hline\hline
total& $1$ & $n+1$ & $2n$ \\  \hline
\end{tabular}
\end{minipage}
\end{table}

The resulting theory in four space-time dimensions exhibits the
same symmetry (\ref{Btrans}) as the original theory in five
dimensions. Besides, a number of extra symmetries emerge that also
find their origin in the five-dimensional theory.
First of all, the extra vector field emerging from the
five-dimensional metric has a corresponding gauge invariance
related to reparametrizations of the extra fifth coordinate by
functions that depend only on the four space-time coordinates. Then
there are special gauge transformations of the $n$ vector fields
with gauge functions that depend exclusively and linearly on the
fifth coordinate. Under these transformations the fifth
component of each gauge field transforms with a constant
translation, whereas the remaining four-dimensional gauge fields
transform linearly into the gauge field originating from
the five-dimensional metric. Finally there are the scale
transformations of the fifth coordinate. The five-dimensional
origin of the duality invariances in four dimensions and their
parameters is concisely summarized by
\begin{eqnarray}
\tilde B^A_{\;B} &\Longrightarrow& \tilde B^A_{\;B}\qquad\qquad
\nonumber \\
\mbox{gauge transformations} \propto x^5  &\Longrightarrow& b^A
\nonumber\\
\mbox{scale transformation of}\; x^5  &\Longrightarrow&  \beta
\nonumber
\end{eqnarray}
Hence altogether we have $n+1$ extra symmetries associated with
the parameters $b^A$ and $\beta$, and $n+1$ extra scalar fields.
In addition to that, there may be {\em hidden symmetries} for
which there exists no explanation in terms of the underlying
higher-dimensional theory. In terms of special coordinates $z^A$
the combined transformations take the form
\begin{equation}
\delta z^A = b^A - \textstyle{2\over 3} \beta \,z^A +\tilde
B^A_{\;B}\, z^B + \textstyle{1\over 2} R^A{}_{\!BC}{}^D\, a_D\, z^B
z^C \,,
\label{ztrans}
\end{equation}
where the possible hidden symmetries are characterized by
the parameters $a_A$. Hidden symmetries exist for those independent
parameters $a_A$ for which $R^A{}_{\!BC}{}^D\,a_D$ is constant
(in special coordinates).
The maximal number of hidden symmetries is thus realized whenever the
curvature tensor itself is constant. In that case the
corresponding K\"ahler space is symmetric \cite{BEC,CremVP}.

The symmetries (\ref{ztrans})
correspond to invariances
of the full supergravity theory in four dimensions and it is
known that they comprise the full isometry group of the K\"ahler
metric for this class of spaces \cite{sssl}.  The root
lattice corresponding to these transformations consists of the
root lattice for the subgroup corresponding to $\tilde B^A_{\;B}$
extended with one dimension associated with the eigenvalue of
the roots under the $\beta$-symmetry. This leads
to a characteristic lattice such as shown on the next page for a
particular example with $n=3$ and
$F(X) = 3 i\left(X^2/X^0\right)\left(X^2X^1 - (X^3)^2\right)$.
In this case the subgroup associated with the parameters
$\tilde B^A_{\;B}$  has two generators denoted by
$\tilde B_2$ and $\tilde B_3$. \hfil Their
roots correspond to the solvable algebra of $SU(1,1)$,\hfil which is
ex-
\newpage

\noindent
\begin{minipage}{7.2cm}  tended to a six-dimensional solvable
algebra by the roots associated with the parameters $\beta$,
$b^1$, $b^2$ and $b^3$. In this case there is one {\em hidden}
symmetry associated with the parameter $a_2$, indicated in the
diagram by $\diamond$.

\setlength{\parindent}{6mm}
The root lattice in the above example reflects the general
situation. Decomposing the full symmetry
algebra $\cal W$ into eigenspaces of the
generator associated with the $\beta$ symmetry  (in the adjoint
representation), we always have \cite{BEC}
\end{minipage}
\hfill
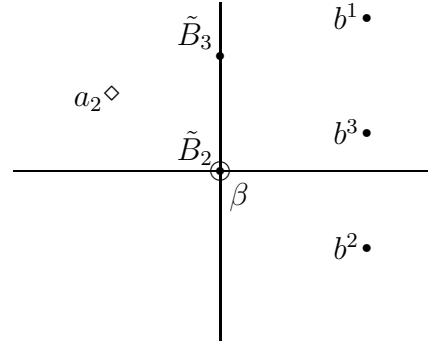
\begin{minipage}{6cm}
\setlength{\unitlength}{0.5mm}
\begin{picture}(120,95)
\put(0,45){\line(1,0){110}}
\put(55,0){\line(0,1){90}}
\put(55,45){\circle*{2}}
\put(55,45){\circle{5}}
\put(55,75.6){\circle*{2}}
\multiput(93.9,24.6)(0,30.6){3}{\circle*{2}}
\put(43.5,48){\makebox(0,0)[bl]{$\tilde B_2$}}
\put(43.5,78.6){\makebox(0,0)[bl]{$\tilde B_3$}}
\put(57.4,36){\makebox(0,0)[bl]{$\beta $}}
\put(85,21.6){\makebox(0,0)[bl]{$b^2$}}
\put(85,52.2){\makebox(0,0)[bl]{$b^3$}}
\put(85,82.8){\makebox(0,0)[bl]{$b^1$}}
\put(26.1,65.4){\makebox(0,0){$\diamond$}}
\put(23.7,62.4){\makebox(0,0)[br]{$a_2$}}
\end{picture}
\end{minipage}

\vspace{3mm}

\begin{equation}
{\cal W}= {\cal W}_{-2/3} +{\cal W}_0 + {\cal W}_{2/3} ,
\label{Kroots}
\end{equation}
where the subscript denotes the eigenvalue with respect to the
$\beta$ symmetry. As it turns out,  ${\cal W}_0$ corresponds to
the subalgebra associated with the parameters $\tilde B^A_{\;B}$
and $\beta$, ${\cal W}_{{2/ 3}}$ contains the generators
corresponding to the
parameters $b^A$, and all possible generators corresponding to
the hidden symmetries belong to ${\cal W}_{-2/3}$. The dimension
of ${\cal W}_{-{2/ 3}}$ is thus at most equal to $n$, whereas the
dimension of ${\cal W}_{{2/ 3}}$ is always equal to $n$.
Unless we have maximal symmetry (i.e. unless there are $n$
independent symmetries associated with the parameters $a_A$, in
which case the space is symmetric) the
isometry group of the corresponding K\"ahler space is not
semi-simple.

\section{Special quaternionic manifolds}
\noindent
We now return to the $N=2$ Maxwell-Einstein Lagrangian
in four dimensions based on a general (holomorphic and
homogeneous) function $F(X)$ and consider
its reduction to three space-time dimensions. In this case an
extra feature is present, because the standard (abelian) gauge-field
Lagrangian in three dimensions can be converted to a scalar-field
Lagrangian by means of a duality transformation. Each
four-dimensional gauge field thus gives rise to two
scalar fields, one of which its component in the fourth dimension and
the other a scalar field resulting from the
conversion of the three-dimensional gauge field. Up to total
divergences only derivatives
of the new scalar field (introduced as a Lagrange multiplier to
impose the Bianchi identity) appear in the Lagrangian, so that in
addition to the original gauge tranformations there is a second
symmetry corresponding to constant shifts of the new field. These
two invariances have parameters denoted by $\alpha^I$ and
$\beta_I$. The same conversion can be carried out for
the vector field that emerges from the three-dimensional metric, so
that the four-dimensional metric gives rise to a
three-dimensional metric and two scalar fields. These scalar
fields are also subject to two invariances, one related to the
scale transformation of the extra coordinate with parameter
$\epsilon^0$ and another one
corresponding to the converted three-dimensional vector field with
parameter $\epsilon^+$.
Altogether, the
Lagrangian (\ref{4dLagr}) thus gives rise to $4(n+1)$ scalar
fields, as shown in table~3,
coupled to gravity with $2n+4$ additional invariances. The
four-dimensional origin of these invariances can be summarized
as follows:
\begin{eqnarray}
\mbox{K\"ahler isometries} &\Longrightarrow& \mbox{K\"ahler
isometries} \nonumber \\
\mbox{gauge transformations} \propto x^4  &\Longrightarrow&
\alpha^I \nonumber\\
\mbox{Lagrange multiplier shifts}  &\Longrightarrow&
\beta_I \ , \ \epsilon^+ \nonumber\\
\mbox{scale transformation of}\; x^4  &\Longrightarrow&
\epsilon^0  \nonumber
\end{eqnarray}

\begin{table}[t]
\begin{minipage}{6cm} Table 3: Decomposition of the $d=4$ bosonic
fields into $d=3$ fields.
\end{minipage} \hfill
\begin{minipage}{7cm}
\begin{tabular}{|r||c|c||}\hline
$d=4$ & metric  & scalars \\ \hline \hline
metric & $1$  & $2$  \\
$n+1$ vectors & $0$ & $2n+2$ \\
$2n$ scalars & $0$ & $2n$ \\ \hline\hline
total& $1$ &  $4n+4$ \\  \hline
\end{tabular}
\end{minipage}
\end{table}

As the Lagrangian is still supersymmetric the scalar
fields must define a quaternionic non-linear sigma model of
quaternionic dimension $n+1$ \cite{dWTNic}. The
corresponding quaternionic spaces are called {\em special}
quaternionic spaces and obviously depend on a homogeneous holomorphic
function of second degree. Like all quaternionic spaces, they are
irreducible Einstein spaces. The explicit Lagrangian was determined in
\cite{Sabi}, where the quaternionic structure was explicitly
verified, and in \cite{dWVP2}, where the complete set of
isometries was determined.

Again we consider the root lattice corresponding to the isometries
of the quaternionic space.  It consists of the
root lattice associated with the invariance of the corresponding special
K\"ahler space extended with one dimension
associated with the eigenvalue of the roots under the scale
symmetry with parameter $\epsilon^0$. The roots of the extra
symmetries noted above take a characteristic position, as can be
shown from the example below, where we have
exhibited the $n=1$ case with $F(X) = i(X^1)^3/X^0$.

\vspace{.8mm}
\noindent
\begin{minipage}{6.3cm} The roots
corresponding to the $SU(1,1)$
isometries of the K\"ahler manifold, denoted
by $b^1$, $\beta$
and $a_1$, are extended with the roots belonging to the extra
transformations that emerge from the
reduction to three dimensions. Observe that there are no
K\"ahler isometries associated with the matrices $\tilde
B^A_{\;B}$ in this case, because the corresponding
five-dimensional theory is {\em pure} supergravity. Just as
before {\em hidden} symmetries emerge, which take
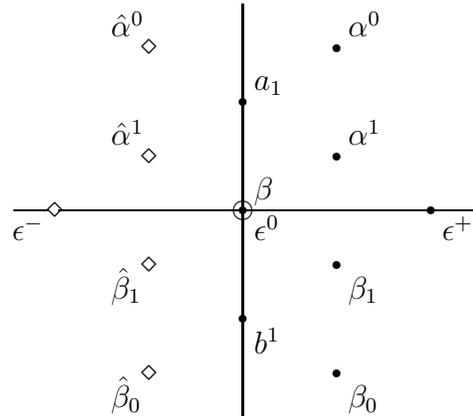
\linebreak
{~}
\end{minipage} \hfill
\begin{minipage}{7cm}
\setlength{\unitlength}{0.5mm}
\begin{picture}(140,115)
\put(4,60){\line(1,0){9}}
\put(16,60){\line(1,0){110}}
\put(65,5){\line(0,1){110}}
\multiput(65,60)(50,0){2}{\circle*{2}}
\put(65,60){\circle{5}}
\put(90,16.7){\circle*{2}}
\put(90,45.6){\circle*{2}}
\put(90,74.4){\circle*{2}}
\put(90,103.3){\circle*{2}}
\multiput(65,31.1)(0,57.8){2}{\circle*{2}}
\multiput(40,16.7)(0,28.9){4}{\makebox(0,0){$\diamond$}}
\put(15,60){\makebox(0,0){$\diamond$}}
\put(12,52){\makebox(0,0)[br]{$\epsilon^-$}}
\put(68,52){\makebox(0,0)[bl]{$\epsilon ^0$}}
\put(118,52){\makebox(0,0)[bl]{$\epsilon ^+$}}
\put(93,7.7){\makebox(0,0)[bl]{$\beta _0$}}
\put(93,36.6){\makebox(0,0)[bl]{$\beta _1$}}
\put(93,77.4){\makebox(0,0)[bl]{$\alpha^1$}}
\put(93,106.3){\makebox(0,0)[bl]{$\alpha^0$}}
\put(68,22){\makebox(0,0)[bl]{$b^1$}}
\put(68,63){\makebox(0,0)[bl]{$\beta$}}
\put(68,92){\makebox(0,0)[bl]{$a_1$}}
\put(30,7.7){\makebox(0,0)[bl]{$\hat\beta _0$}}
\put(30,36.6){\makebox(0,0)[bl]{$\hat\beta _1$}}
\put(30,77.4){\makebox(0,0)[bl]{$\hat\alpha^1$}}
\put(30,106.3){\makebox(0,0)[bl]{$\hat\alpha^0$}}
\end{picture}
\end{minipage}

\newpage

\noindent
characteristic positions in the root lattice on the left
half-plane. In the case at hand, there are five such symmetries
indicated by $\diamond$, which extend this diagram to the root
lattice of $G_{2(+2)}$. Its
solvable subalgebra consists of the solvable subalgebra of $SU(1,
1)$, associated with the parameters
$\beta$ and $b^1$, extended by the generators of the six extra
symmetries corresponding to $\epsilon^0$, $\epsilon^+$,
$\alpha^I$ and $\beta_I$.

As shown in \cite{dWVP2} this example is completely
characteristic for the general case. The hidden
symmetries are always associated to roots with eigenvalues
$-1$ or $-\half$ under the $\epsilon^0$ symmetry. The generators
corresponding to  all isometries of the special quaternionic
space decompose generally according to
\begin{equation}
{\cal V}= {\cal V}_{-1} +{\cal V}_{-1/2}
+ {\cal V}_0 +{\cal V}_{1/2} + {\cal V}_1 .  \label{Qroots}
\end{equation}
As shown in \cite{dWVP2}, for symmetric spaces, the dimension of
${\cal V}_{-1}$ and ${\cal V}_{-1/2}$ is equal to 1 and $2n+2$,
respectively. Otherwise ${\cal V}_{-1}$ is empty and the
dimension of ${\cal V}_{-1/2}$ is less
than or equal to $n+1$. The dimension of ${\cal V}_{1/2}$ and
${\cal V}_1$ is always equal to $2n+2$ and 1, respectively. The
parameter associated with the new
symmetry in ${\cal V}_{-1}$ is denoted by $\epsilon^-$;  those
corresponding to the symmetries in
${\cal V}_{-1/2}$ are denoted by $\hat\alpha^I$ and $\hat\beta_I$.
 The maximal number of symmetries exist if and only if
${\cal V}_{-1}$ is not empty. In that case both
the quaternionic manifold and the corresponding  K\"ahler manifold
are symmetric.

\section{Homogeneous quaternionic spaces}
\noindent
As shown above the  three types of special manifolds
are related by dimensional reduction of the corresponding
supergravity models. A special real $(n-1)$-dimensional manifold
$\Rbar_{n-1}$  is thus related to a special K\"ahler manifold
$\Cbar_n$ of complex dimension $n$.
Likewise, a special K\"ahler manifold of complex dimension $n$ is
related to a special quaternionic
manifold $\Hbar_{n+1}$ of quaternionic dimension $n+1$. These
relations define the so-called $\bf r$ map from special real to
special K\"ahler manifolds, and the $\bf c$ map from special
K\"ahler to special quaternionic manifolds. Under both maps the
isometry groups are enlarged and their rank is increased by
precisely one unit. In fact one can
show that the number of additional symmetries is at least as
large as the number of new coordinates that emerge under the
map.  It should be clear that the inverse $\bf r$  and $\bf c$ maps
do not always exist. For instance, only the K\"ahler manifolds
based on the functions (\ref{Fd}) (and the functions that lead to
equivalent field equations) can be in the image of
the $\bf r$ map.

Another important property of the maps is that the homogeneity of
the manifold is preserved, provided there is a transitive
subgroup of the isometry group that can be extended to a full
invariance group of the corresponding supergravity theory.
Conversely, if a homogeneous manifold is in the image of one of
the maps, then the original manifold must also be homogeneous
\cite{dWVPVan}. Therefore the maps have particular relevance for
homogeneous special spaces. Actually, already in \cite{CecFerGir}
it was recognized that \Al's classification of {\em normal}
quaternionic spaces \cite{Aleks} amounts to the construction of
an inverse $\bf c$ map.
Normal quaternionic spaces are quaternionic spaces that admit a transitive
completely solvable group of motions. It was conjectured in
\cite{Aleks} that the homogeneous quaternionic spaces consist of
compact symmetric quaternionic and
normal quaternionic spaces. The algebra corresponding to the
group of solvable motions for the normal spaces can be decomposed
according to the eigenvalues of one of its generators $e_0$. This
generator can be identified with the generator associated with
the $\epsilon^0$ symmetry of the previous section, and we find
that the solvable algebra can be embedded in ${\cal V}_0 +{\cal
V}_{1/2}+{\cal V}_1$ (cf. (\ref{Qroots})).

According to \Al\ there are two different
types of normal quaternionic spaces characterized by their
so-called canonical quaternionic subalgebra. The first type with
subalgebra $C_1^1$ turns out to correspond to the quaternionic
projective spaces $USp(2n+2,2)/(USp(2n+2)\otimes SU(2))$.  Their {\em
solvable} algebra contains only one element in ${\cal V}_0$
(namely the generator $e_0$), and $4n$ and 3 elements in ${\cal
V}_{1/2}$ and ${\cal V}_1$, respectively. However, from section~4
we know that the
special quaternionic spaces contain only one generator in ${\cal
V}_1$, so that the quaternionic projective spaces are {\em not}
special. Their rank is equal to 1 or 0 (the latter corresponds to
the empty space). Although they do not appear in the image of the
$\bf c$ map, they can be coupled to $N=2$ supergravity in three
or four dimensions. The rank-0 case corresponds to pure
supergravity, as is indicated in table~\ref{homnonsp}.

The second type of normal spaces has a canonical subalgebra
$A_1^1$. The structure of the solvable algebra is now as
follows:  ${\cal V}_0$ contains the direct sum of $e_0$ and a normal
K\"ahler algebra ${\cal W}^{\rm s}$ of dimension $2n$, ${\cal
V}_{1/2}$ contains $2n+2$ generators and ${\cal V}_1$ contains
precisely 1 generator. In order to be quaternionic, the
representation of ${\cal W}^{\rm s}$ induced by the adjoint
representation of the solvable algebra on the $2n+2$ generators
in ${\cal V}_{1/2}$ must generate a
solvable subgroup of $Sp(2n+2, \Rbar)$. Therefore each normal
quaternionic space of this type defines the basic ingredients of
a special normal K\"ahler space, encoded in its solvable
transitive group of K\"ahler isometries.
\Al's analysis thus strongly indicates that a corresponding
four-dimensional $N\!=\!2$ supergravity theory should exist, so
that under the $\bf c$ map one will recover the original normal
quaternionic space. To establish the existence of  the
supergravity theory, one must prove that a corresponding
holomorphic function $F(X)$ exists that allows for these K\"ahler
isometries. This
program was carried out by Cecotti \cite{Cecotti}, who explicitly
constructed the function $F(X)$ corresponding to each of the
normal quaternionic spaces with canonical subalgebra $A^1_1$. For
rank 2 one has the so-called minimal coupling, where $F(X)$ is a
quadratic polynomial, and a special case corresponding to $F(X)=
i(X^1)^3/X^0$. This is the example discussed in section~4. The
corresponding K\"ahler space is the image under the {\bf r} map
of the empty real space. In other words, it is the space one
obtains after reducing
pure $d=5$ supergravity to four dimensions (see
table~\ref{homnonsp}). The surprising feature of Cecotti's result
is that all remaining normal quaternionic spaces (which are of
rank 3 or 4) are indeed in the image of the {\bf c} map and are
associated with functions $F(X)$ that can be brought into the form
(\ref{Fd}). These spaces are therefore also in the image of the
{\bf c}$\scriptstyle\circ${\bf r} map. If \Al's classification is
complete, there can be no other special real or K\"ahler
spaces with solvable transitive groups of isometries that can be
promoted to an invariance group of the corresponding supergravity
Lagrangian.

\setcounter{table}{3}
\begin{table}[tb]
\begin{center}
\begin{tabular}{||c|c|c||c|c||}
\hline
real & K\"ahler & quaternionic & $n\!+\!1$& $R$ \\
\hline &&&&\\[-3mm]
&   &   SG     &$0$&0\\[2mm]
&   & $\frac{USp(2n+2,2)}{USp(2n+2)\otimes SU(2)} $
    &$n+1\geq 0$&1 \\[2mm]
&  SG       &$\frac{U(1,2)}{U(1)\otimes U(2)} $
    & 1&1\\[2mm]
&$\frac{U(n,1)}{U(n)\otimes U(1)}$
    &$\frac{U(n+1,2)}{U(n+1)\otimes U(2)} $ & $n+1\geq 2$
    &2    \\[2mm]
SG  & $\frac{SU(1,1)}{U(1)}$
    &$\frac{G_{2(+2)}}{SU(2)\otimes SU(2)}$ &2 &2\\[2mm]
\hline
\end{tabular}
\end{center}
\caption{Normal quaternionic spaces with rank $R\leq 2$ and
quaternionic dimension $n\!+\!1$ and the corresponding special
real and K\"ahler spaces (whenever they exist). "SG" means pure
supergravity and thus corresponds to the empty space.  }
\label{homnonsp}
\end{table}

This result suggests an alternative approach. Namely one can start
from the real special spaces, and try to classify all the
homogeneous spaces with an invariance group (\ref{Btrans}) that
acts
transitively on the sigma-model manifold. This classification was
carried out in \cite{dWVP3}. Subsequently one can apply the $\bf
r$ and the $\bf c$ map to obtain all the corresponding
homogeneous K\"ahler and quaternionic spaces. These spaces can
then be confronted with those in the classification of
\cite{Aleks,Cecotti} with rank 3 or 4.

Let us therefore briefly describe the classification of the
homogeneous real spaces. First one introduces the so-called
canonical parametrization of the coefficients $d_{ABC}$ by choosing a
reference point on the manifold within the domain where the
kinetic terms for the various fields of the
corresponding supergravity Lagrangian have the proper signs
\cite{BEC,GuSiTo}. After suitable reparametrizations this
reference point equals $h^A = (1 ,0, \ldots,0)$ and the
$d$-coefficients are restricted to $d_{111}=1$, $d_{11a} =0$, $d_{1ab}
=-\textstyle{\frac{1}{2}} \delta_{ab}$, with $d_{abc}$
unrestricted ($a,b = 2,\ldots , n$). This parametrization is preserved
under rotations of the fields $h^2,\ldots,h^n$, which leave the reference
point invariant. A subgroup of these orthogonal transformations
that leaves $d_{abc}$ invariant thus defines the
isotropy group of the homogeneous space, while the isometry
group is the invariance group of the
$d_{ABC}$ tensor. Therefore the homogeneous space is just the coset
space of these two groups. The condition that the isometries act
transitively on the reference point takes the form
\begin{equation}
 d_{e(ab}\,d_{cd)e}-{\textstyle\half} \,
\delta_{(ab}\,\delta_{cd)}\,  =  d_{e(ab}\, A_{c)e;d}\ ,
\label{homeq}
\end{equation}
where $A_{ab;c}$ is an arbitrary tensor antisymmetric in the
first two indices. The homogeneous equation (i.e., with
$A_{ab;c}=0$) was already solved in \cite{BEC}. According to
\cite{dWVP3} the solution of
the inhomogeneous equation can be expressed as follows.
First we decompose the indices $A$ into $A= 1, 2, \mu, i$,  with
$\mu=1,\ldots, q+1$ and $i=1,\ldots,r$, so that $n= 3+q+r$. Hence
we assume $n\geq 2$. We then establish that the general solution
of (\ref{homeq}) takes the form (after some redefinitions, so
that we are no longer in the canonical parametrization)
\begin{equation}
d_{ABC}\,h^Ah^Bh^C = 3\Big\{ h^1\,\big(h^2\big)^2 -h^1\,
\big(h^\mu\big)^2 -h^2\,\big(h^i\big)^2
+\gamma_{\mu ij}\,h^\mu\, h^i\,h^j \Big\}\ .\label{genFd}
\end{equation}
Here the coefficients $\gamma_{\mu ij}$ are $r\times r$ gamma
matrices that generate a real
$(q\!+\!1)$-dimensional Clifford algebra of positive signature. This
property severely constrains the possible values for $q$ and $r$.
The gamma matrices do not necessarily correspond to irreducible
representations of the Clifford algebra and are thus uniquely
specified (up to similarity transformations) by the
multiplicity $P$ of the irreducible representations contained in
them. However, when $q$ is a multiple of 4, there exist two
inequivalent irreducible representations and we have to specify
two corresponding multiplicities, $P$ and $\dot P$. The
polynomials (\ref{genFd}) and their corresponding
manifolds are denoted by by $L(q,P)$ or $L(4m,P,\dot P)$,
depending on whether there exist inequivalent Clifford algebra
representations.

Most of the isometries\footnote{There may be additional
isometries of the special real space, but those cannot be promoted to
an invariance of the full corresponding supergravity theory
\cite{sssl}}
correspond to invariance tansformations of (\ref{genFd}). They
contain three obvious subgroups. There are the scale
transformations
\begin{equation}
h^1\to e^{2\lambda} h^1\ , \quad (h^2, h^\mu) \to e^{-\lambda} (h^2,
h^\mu) \, , \quad h^i\to e^{{1\over 2} \lambda} h^i \,,
\label{scale}
\end{equation}
the $SO(q+1)$ rotations that act on $h^\mu$ in the vector and on
$h^i$ in the spinor representation, and finally the
metric-preserving elements in the centralizer of the Clifford
algebra, which act exclusively on $h^i$. The latter constitute a
subgroup ${\cal S}_q(P,\dot P)$, which equals $SO(P)\otimes
SO(\dot P)$, $U(P)\otimes U(\dot P)$ or $USp(P)\otimes USp(\dot P)$,
depending on the value for $q$.

Even in the generic case, there are additional symmetries
present. The corresponding algebra decomposes into eigenspaces of
the generator associated with the scale transformations
(\ref{scale}) according to
\begin{equation}
{\cal X}={\cal X}_{-3/2} + {\cal X}_0+{\cal X}_{3/2}\,.
\label{Rroots}
\end{equation}
Here ${\cal X}_0$ contains the generators of three groups
mentioned above and
$q+1$ extra generators. The latter extend the algebra of $SO(q+1)$
to that of $SO(q+1,1)$, so that ${\cal X}_0$ contains the
generators belonging to $SO(1,1)\otimes SO(q+1,1)\otimes {\cal
S}_q(P,\dot P)$, where $SO(1,1)$ denotes the scale
transformations (\ref{scale}). It thus follows that the solvable
algebra contained in ${\cal X}_0$ has dimension $q+2$ with rank
2, except for $q=-1$ when the rank equals 1.
The $r$ generators in ${\cal X}_{3/2}$ are present for all
cases and extend the dimension of the solvable subalgebra to
$q+2+r= n-1$, which equals the dimension of the corresponding
real space.

Furthermore in four special cases
there are yet another $r$ generators contained in ${\cal
X}_{-3/2}$. These four cases are characterized by ${1\over 2}r=q=
1,2,4$ or 8, so that $n= 6,9,15$ or 27, respectively. (We remind
the reader that the dimension of the real space equals $n-1$.)
The corresponding manifolds are symmetric and related to Jordan algebras
and the magic square \cite{GuSiTo}. Their images under the {\bf
r} and {\bf c} map are shown in table~5. Hence we are dealing
with a root lattice for $\cal X$ that is rather similar to the
root lattices encountered previously for the K\"ahler and
quaternionic spaces. As an example con-

\vspace{.9mm}\noindent
\begin{minipage}{6cm}
sider the root lattice corresponding to $L(1,1)$,
shown right. The roots on the vertical axis belong to $SO(1,
1)\otimes SO(2,1)$. The two roots denoted by $\diamond$
exist only for the special case of $L(1,1)$ (and likewise for
$L(2,1)$, $L(4,1)$ and $L(8,1)$, where we have 4, 8 and 16 such
roots, respectively), unlike the other
roots which exist for all generic homogeneous special spaces.
\end{minipage}
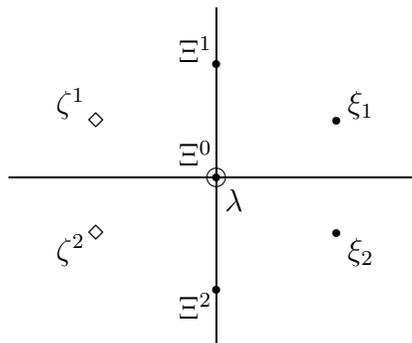
\hfill
\begin{minipage}{7cm}
\setlength{\unitlength}{0.5mm}
\begin{picture}(120,110)
\put(0,50){\line(1,0){110}}
\put(55,5){\line(0,1){90}}
\put(55,50){\circle*{2}}
\put(55,20){\circle*{2}}
\put(55,50){\circle{5}}
\put(55,80){\circle*{2}}
\multiput(87,35)(0,30){2}{\circle*{2}}
\put(45,53){\makebox(0,0)[bl]{$\Xi^0$}}
\put(45,80.6){\makebox(0,0)[bl]{$\Xi^1$}}
\put(45,12,4){\makebox(0,0)[bl]{$\Xi^2$}}
\put(57.4,41){\makebox(0,0)[bl]{$\lambda$}}
\put(90,28){\makebox(0,0)[bl]{$\xi_2$}}
\put(90,67){\makebox(0,0)[bl]{$\xi_1$}}
\put(20,67){\makebox(0,0)[br]{$\zeta^1$}}
\multiput(23,35)(0,30){2}{\makebox(0,0){$\diamond$}}
\put(20,28){\makebox(0,0)[br]{$\zeta^2$}}
\end{picture}
\end{minipage}

\vspace{1mm}
Hence we find the complete set of homogeneous real spaces, which
are of rank 1 or 2 and shown in table~5. Upon application of the
{\bf r} and {\bf c}$\scriptstyle\circ${\bf r} maps they give rise
to K\"ahler and quaternionic spaces of rank 2 or 3 and 3 or 4,
respectively. Continuing this analysis one can systematically
determine the isometry and isotropy groups of these spaces
\cite{dWVPVan}. When comparing the results with the
classification of \cite{Aleks,Cecotti} one finds that there exist
additional homogeneous spaces, which in table~5 have been
indicated by a $\star$.

\begin{table}[t]
\begin{center}
\begin{tabular}{||l|ccc|c||}\hline
$C(h)$&real & K\"ahler & quaternionic   \\
\hline&&&\\[-3mm]
$L(-1,0)$&$SO(1,1)$&$\left[\frac{SU(1,1)}{U(1)}\right]^2$&$\frac{SO(3,4)}{(
S U ( 2 ) ) ^ 3 } $ \\[2mm]
$L(-1,P)$&$\frac{SO(P+1,1)}{SO(P+1)}$& $\star$ & $\star$ \\[2mm]
\hline&&&\\[-3mm]
$L(0,0)$&$[SO(1,1)]^2$&$\left[ \frac{SU(1,1)}{U(1)}\right] ^3$&$
\frac{SO(4,4)}{SO(4)\otimes SO(4)} $\\[2mm]
$L(0,P)$&$\frac{SO(P+1,1)}{SO(P+1)}\otimes
SO(1,1)$&$\frac{SU(1,1)}{U(1)}\otimes
\frac{SO(P+2,2)}{SO(P+2)\otimes SO(2)}$&$
\frac{SO(P+4,4)}{SO(P+4)\otimes SO(4)} $\\[2mm]
$L(0,P,\dot P)$&$Y(P,\dot P)$&$K(P,\dot P)$&$W(P,\dot P)$ \\[2mm]
$L(q,P)$&$X(P,q)$&$H(P,q)$&$V(P,q)$\\[2mm]
$L(4m,P,\dot P)$& $\star$ & $\star$& $\star$\\[2mm]
$L(1,1)$&
$\frac{S\ell(3,\Rbar)}{SO(3)}$&$\frac{Sp(6)}{U(3)
}$&$\frac{F_4}{USp(6)\otimes SU(2)}$\\[2mm]
$L(2,1)$&
$\frac{S\ell(3,\Cbar)}{SU(3)}$&$\frac{SU(3,3)}{SU(3)\otimes
SU(3)\otimes U(1)}$&$\frac{E_6}{SU(6)\otimes SU(2)}$\\[2mm]
$L(4,1)$&
$\frac{SU^*(6)}{Sp(3)}$&$\frac{SO^*(12)}{SU(6)\otimes
U(1)}$&$\frac{E_7}{\overline{SO(12)}\otimes SU(2)}$\\[2mm]
$L(8,1)$&
$\frac{E_6}{F_4}$&$\frac{E_7}{E_6\otimes
 U(1)}$&$\frac{E_8}{E_7\otimes SU(2)}$\\[2mm]
\hline \end{tabular}
\end{center}
\caption{Homogeneous special real spaces and their corresponding
K\"ahler and quaternionic spaces. The rank of the real spaces is
equal to 1 (above the line) or 2 (below the line). The rank of
the corresponding K\"ahler and quaternionic manifolds is increased by
one or two units, respectively; $P$, $\dot P$, $q$
and $m$ are arbitrary positive integers. } \label{homsp}
\end{table}

We close with a few comments on the spaces denoted by $L(-1,P)$.
The real spaces correspond to
\begin{equation}
L(-1,P)\,:\quad {SO(P+1,1)\over SO(P+1)}\ . \qquad (n= P+2)
\end{equation}
They were constructed in \cite{GuSiTo2}, but the corresponding
K\"ahler spaces that emerge under the $\bf r$ map were not
correctly identified. Moreover, the corresponding special K\"ahler
and quaternionic spaces are not contained in the classifications
of \cite{Aleks,Cecotti}. The rank of these spaces equals 1, 2 and
3, for the real, K\"ahler and quaternionic versions,
respectively. Only
the case $P=0$, corresponding to a $n=2$ symmetric real, K\"ahler
and quaternionic space, appeared in the classifications (see
table~5). In spite of the fact that the  real
spaces are symmetric, the corresponding K\"ahler and quaternionic
spaces are not symmetric for $P>0$ . The reason for the extra
symmetry of the real sigma models is that they have extra
invariances that are not preserved by the interactions with the
vector fields in $d=5$ supergravity \cite{sssl}. Therefore these
isometries are not preserved under the $\bf r$ map. Observe that
the root lattice for the real and the K\"ahler spaces
corresponding to $L(-1,1)$ was already discussed in section~3.
For further details and an extensive list of references we refer
to \cite{dWVPVan}.

\vspace{4mm}\noindent
This work was carried out in the framework of the European
Community Research Programme ``Gauge theories, applied
supersymmetry and quantum gravity", with a financial contribution
under contract SC1-CT92-0789.


\begin{thebibliography}{99}
\bibitem{dWVP} B. de Wit and A. Van Proeyen, Nucl. Phys. {\bf B245} (1984) 89.
\bibitem{special} A.~Strominger, Commun.~Math.~Phys. {\bf 133}
(1990) 163.
\bibitem{GrosCand} D.~Gross, J.~Harvey, E.~Martinec and  R.~Rohm,
Nucl. Phys. {\bf B256} (1985) 253, {\bf B267} (1986) 75. \\
P.~Candelas, G.T.~Horowitz, A.~Strominger and E.~Witten, Nucl.
Phys. {\bf B258} (1985) 46.
\bibitem{Seiberg} N. Seiberg, Nucl. Phys. {\bf B303} (1988) 286.
\bibitem{Cand} P.~Candelas and X.~C.~de~la~Ossa, Nucl. Phys. {\bf B355}
(1991) 455,\\
P.~Candelas, X.~C.~de~la~Ossa, P.~Green and
L.~Parkes, Phys.~Lett. {\bf 258B} (1991) 118; Nucl.~Phys. {\bf
B359} (1991) 21.
\bibitem{BoCa} M. Bodner and A.C. Cadavid, Class. Quantum Grav. {\bf 7}
(1990) 829,\\
M.~Bodner, A.C.~Cadavid and S.~Ferrara, Class. Quantum Grav. {\bf
8} (1991) 789.
\bibitem{mirror} L.J.~Dixon, in {\it Superstrings, Unified
Theories and Cosmology 1987}, eds.
G.~Furlan et al. (World Scientific, 1988), p. 67. \\
W.~Lerche, C.~Vafa and N.P.~Warner, Nucl. Phys. {\bf B324} (1989)
427, \\
B.R.~Greene and M.R.~Plesser, Nucl. Phys. {\bf B338} (1990) 15,\\
P.~Candelas, M.~Lynker and R.~Schimmrigk, Nucl. Phys. {\bf B341}
(1990) 383.
\bibitem{CecFerGir} S. Cecotti, S. Ferrara and L. Girardello,
Int. J. Mod. Phys. {\bf A4} (1989) 2457.
\bibitem{GuSiTo} M.~G\"unaydin, G.~Sierra and P.K.~Townsend,
Phys.~Lett. {\bf 133B} (1983) 72;
Nucl.~Phys. {\bf B242} (1984) 244, {\bf B253} (1985) 573.
\bibitem{BagWit} J.~Bagger and E.~Witten, Nucl.~Phys. {\bf B222}
(1983) 1.
\bibitem{dWTNic} B.~de~Wit, A.~Tollst\'en and H.~Nicolai,
Nucl. Phys. {\bf B392} (1993) 3.
\bibitem{Aleks} D.V. \Al , Math. USSR Izvestija {\bf 9} (1975) 297.
\bibitem{Cecotti} S. Cecotti, Commun. Math. Phys. {\bf 124} (1989) 23.
\bibitem{dWVPVan} B.~de~Wit, F.~Vanderseypen and A.~Van~Proeyen,
Nucl. Phys. {\bf B400} (1993) 463.
\bibitem{BEC} E. Cremmer, C. Kounnas, A. Van Proeyen, J.P. Derendinger, S.
Ferrara, B. de Wit and L. Girardello, Nucl. Phys. {\bf B250} (1985) 385.
\bibitem{CremVP} E. Cremmer and A. Van Proeyen, Class. Quantum Grav. {\bf 2}
(1985) 445.
\bibitem{sssl} B. de Wit and A. Van Proeyen, Phys.~Lett. {\bf
B293} (1992) 94.
\bibitem{Sabi} S. Ferrara and S. Sabharwal, Nucl. Phys. {\bf B332} (1990)
317.
\bibitem{dWVP2} B.~de~Wit and A.~Van~Proeyen, Phys.~Lett. {\bf
B252} (1990) 221.
\bibitem{dWVP3}  B.~de~Wit and A.~Van~Proeyen,
Commun.~Math.~Phys. {\bf 149} (1992) 307.
\bibitem{GuSiTo2} M.~G\"unaydin, G.~Sierra and P.K.~Townsend,
Class.~Quantum~Grav. {\bf 3} (1986) 763.
\end{thebibliography}
\end{document}